%% file: main.tex
\documentclass[11pt]{article}

\usepackage{fontawesome5}

\usepackage[margin=1in]{geometry}
\usepackage{times}
\usepackage{natbib}
\usepackage{latexsym}

\usepackage{dblfloatfix}
\usepackage[T1]{fontenc}
\usepackage[utf8]{inputenc}

\usepackage{microtype}

\AtBeginDocument{\setlength{\parfillskip}{0pt plus 0.58\textwidth}}

\usepackage{inconsolata}
\usepackage[dvipsnames]{xcolor}

\usepackage{graphicx}
\graphicspath{{figures/}}
\usepackage[colorlinks,linkcolor=orange, anchorcolor=blue,citecolor=blue,urlcolor=MidnightBlue]{hyperref}   
\usepackage{url}

\usepackage{multirow}
\usepackage{amsfonts,amsthm,amsmath,amssymb}
\usepackage{enumitem}

\usepackage{subcaption}
\usepackage{float}
\usepackage{longtable}

\theoremstyle{remark}
\newtheorem{remark}{Remark}

\newcommand{\algname}{\texttt{Anchor-ECC}}

\title{Anchor-ECC: Local Integrity Checking for Watermarked LLM Outputs via Error-Correcting Codes}

\author{%
Zewei Deng$^{1}$, Muhammad Siddeek$^{2}$, Liyan Xie$^{1}$, Mohamed Seif$^{3,5}$\\
Mengdi Wang$^{5}$, H. Vincent Poor$^{5}$, Andrea Goldsmith$^{4,5}$\\[0.55em]
\small $^{1}$University of Minnesota \quad $^{2}$Google \quad
$^{3}$Oakland University\\
\small $^{4}$Stony Brook University \quad
$^{5}$Princeton University\\[0.35em]
\small \href{https://zeweid1221.github.io/Anchor-ECC-Website/}{\faGlobe\; Project Page}
\quad$\vert$\quad
\href{https://github.com/zeweid1221/ECC_Watermark}{\faGithub\; Code}
}
\date{}

\begin{document}
\maketitle

\begin{abstract}
LLM watermarking has become an effective approach to distinguishing AI-generated text from human-written text by embedding detectable patterns during generation. However, a small post-generation edit may change the meaning of the text without removing its overall watermark signal, creating a risk that the modified content is still attributed to the original model. We propose \algname{}, which incorporates the error-correcting code (ECC) constraints and explicit boundary anchors into the watermark structure and pairs them with a dynamic-programming decoder to detect and localize post-generation edits. Across Qwen3-8B, Mistral-7B-Instruct-v0.3, and OPT-125M, the approximate-hard setting achieves about $99.7\%$ block-level true positive rate (TPR) with at most $7.6\%$ false alarm rate (FAR) for edit detection under mixed insertions, deletions, and substitutions, while preserving the distinction between watermarked outputs and unwatermarked text. Additional quality experiments identify lower-perplexity configurations that retain strong edit-detection performance. Together, these results extend LLM watermarking from source identification to \emph{local} integrity verification while supporting configurable trade-offs between detection reliability and generation quality.

\end{abstract}

\section{Introduction}

As large language models (LLMs) %such as GPT \cite{kasneci2023chatgpt} and Gemini \cite{team2023gemini} 
become increasingly widespread, the need for techniques to verify content authenticity, ownership, and responsible usage has become increasingly important. Watermarking addresses this need by embedding hidden signals into LLM outputs that enable content tracking, attribution, and verification, helping prevent misuse and ensure accountability \citep{aaronson2023watermarking,dathathri2024scalable,chao2024watermarking}.

% Large language models (LLMs) are increasingly deployed in real-world settings, raising urgent questions about content attribution, authenticity, and misuse prevention.
% Watermarking has been proposed as a principled solution for identifying model-generated text by embedding hidden structural signals during generation.
% While existing watermarking schemes achieve reliable document-level detection \cite{zhao2025efficiently,pan2025waterseeker,li2024segmenting}, they remain vulnerable to post-generation edits.
% Even a small number of token insertions, deletions, or substitutions can disrupt watermark structure without substantially altering semantic content.

Existing watermarking methods and their detectors are largely designed for document- or segment-level decisions, focusing on whether text is generated by an LLM or a human author \citep{zhao2025efficiently,pan2025waterseeker,li2024segmenting}. They offer limited evidence for determining whether and where a watermarked passage was modified after generation. This distinction matters because a local insertion, deletion, or substitution can spoof a claim, stance, certainty level, or source while leaving the document-level watermark detectable. The modified statement may therefore remain associated with the original generator, creating an attribution risk for the content creator \citep{pang2024no,zhou2024bileve}.

In this work, we propose \algname{} for \emph{local integrity checking} of watermarked LLM outputs. It combines two complementary error-correcting-code (ECC) constraints: $(1)$ Varshamov--Tenengolts (VT) constraints handle insertions and deletions \citep{lcvenshtcin1966binary,schoeny2017codes}, while $(2)$ Hamming-type constraints detect substitutions \citep{hamming1950error}. Each joint codeword forms the payload of a short block, followed by a dedicated boundary anchor that separates it from the next block. We realize these structural symbols by partitioning the model vocabulary into two payload buckets for bits 0 and 1 and a disjoint boundary-anchor bucket. At detection time, a \emph{global} decoder reconstructs a minimum-cost block alignment, flags blocks that violate the joint constraints, and narrows each alarm to candidate edit locations. These \emph{localized} alarms direct semantic or human review to short passages that may contain spoofing edits.

The repeated ECC structure also retains the conventional watermark function: a global watermark score can test whether a candidate text is consistent with the keyed watermark. To reduce generation distortion, we further construct semantically balanced payload buckets, select lexically usable boundary anchors with light semantic content, apply finite logit biasing, and complete each codeword adaptively during sampling. These generation choices expose a measurable quality--reliability trade-off while leaving the codebook and detector unchanged.
%To further distinguish structural deviations from semantically meaningful edits, we incorporate a lightweight white-box language model to perform post-hoc semantic intent verification on suspicious regions. 
The main contributions of this paper are summarized as follows:
\begin{itemize}[leftmargin=*, itemsep=0pt, topsep=2pt]
    \item We introduce \algname{}, a blockwise integrity layer that embeds joint VT--Hamming constraints in LLM outputs, turning a document-level watermark into local structural evidence.
    \item We design a global dynamic-programming decoder that recovers latent boundaries, flags inconsistent blocks, and refines each alarm to a set of candidate edit sites.
    \item We evaluate the detection performance of \algname{} against matched baselines across a range of watermarking strengths, using both synthetic and question-aware benign and spoofing edits. We further characterize watermark identifiability and the current realization cost.
    % \item We introduce a semantic intent verification mechanism that refines structural edit detection using contextual likelihood scoring.
\end{itemize}

\subsection{Related Work}
% \noindent \textbf{Watermarking Methods.} 

Our work builds on the green-list watermarking framework, which promotes a keyed subset of tokens by shifting model logits \citep{kirchenbauer2023watermark}; \citet{zhaoprovable} replace the context-dependent green list with a fixed partition and establish robustness guarantees. We are further related to the pattern-based watermarking approach in \citet{chen2025a}, which targets order-agnostic LLMs. There also exists a watermarking scheme that utilizes pseudorandom error-correcting codes \citep{christ2024pseudorandom}, but it focuses on robust watermark detection rather than edit localization. 
Combinatorial watermarking embeds prescribed bucket patterns and flags post-generation edits when those patterns are disrupted \citep{LLM-edit-detect2025}. It targets the same edit-detection task through local sequence structure, making its two- and four-bucket variants matched non-ECC baselines in Section~\ref{sec:numerical}. A synchronization-assisted VT construction combines a synchronization string with VT blocks to detect and localize insertions and deletions \citep{deng2026detecting}. \algname{} instead uses explicit boundary anchors, intersects VT and Hamming constraints to cover substitutions, and recovers corrupted boundaries through a global block-decomposition decoder.

%Our work builds on and is thus close to the provably robust watermarking scheme \citep{kirchenbauer2023watermark}, which perturbs the model’s logit vector in a green list. Common choices of the green list include the KGW scheme \citep{kirchenbauer2023watermark} and the Unigram scheme \citep{zhaoprovable}. 
%
%Our work is also related to \cite{chen2025a}, which proposes a similar pattern-based watermarking but for order-agnostic LLMs. Our work is also closely related to \citet{LLM-edit-detect2025}, a recent effort on detecting post-generation edits to LLM outputs, but differs in that we embed Varshamov–Tenengolts (VT) codes as structured strings within watermarked text and explicitly leverage their insertion–deletion correction properties for more systematic and accurate edit detection.

% \vspace{0.05in}
% \noindent \textbf{Watermarked Segments Detection.} 

There is also a line of work that proposes detectors to flag \textit{long} content, such as AI-generated content detection \citep{bao2023fast,chakraborty2023possibilities,gehrmann2019gltr,li2024robust,mitchell2023detectgpt,sadasivan2023can}. The watermark \emph{agnostic} approach in \citet{kashtan2023information} seeks finer granularity by applying the Higher Criticism metric to detect sparse anomalies. \citet{leipald} introduced a Bayesian framework to flag the LLM-generated segments \citep{cohen2022bayesian}. While their objective is related to ours, they do not consider post-generation edits made to LLM output. 
%Methodologically, their approach is also fundamentally different---they operate on fixed segmentations and do not leverage embedded watermarks. 
Recent work also studies identifying watermarked spans within mixed-source documents. These approaches, such as \citet{zhao2025efficiently} and \citet{pan2025waterseeker}, are designed to detect long, contiguous watermarked regions and assume that the underlying watermarked text remains largely intact. In contrast, we focus on localizing structural changes caused by token-level edits to watermarked LLM output.

\section{Proposed Watermarking and Edit Detection Methods}\label{sec:method}

Let $\mathcal{V}$ denote the tokenizer vocabulary and $x=(x_1,\ldots,x_T)$ a watermarked token sequence. Under \algname{}, each generated token maps to a structural symbol in $\{0,1,2\}$: $0$ and $1$ denote the two payload buckets, while $2$ denotes the boundary-anchor bucket. Post-generation insertions, deletions, and substitutions transform $x$ into an observed sequence $y=(x'_1,\ldots,x'_{T'})$.

\begin{figure*}[t!]
    \centering
    \includegraphics[width=0.99\linewidth]{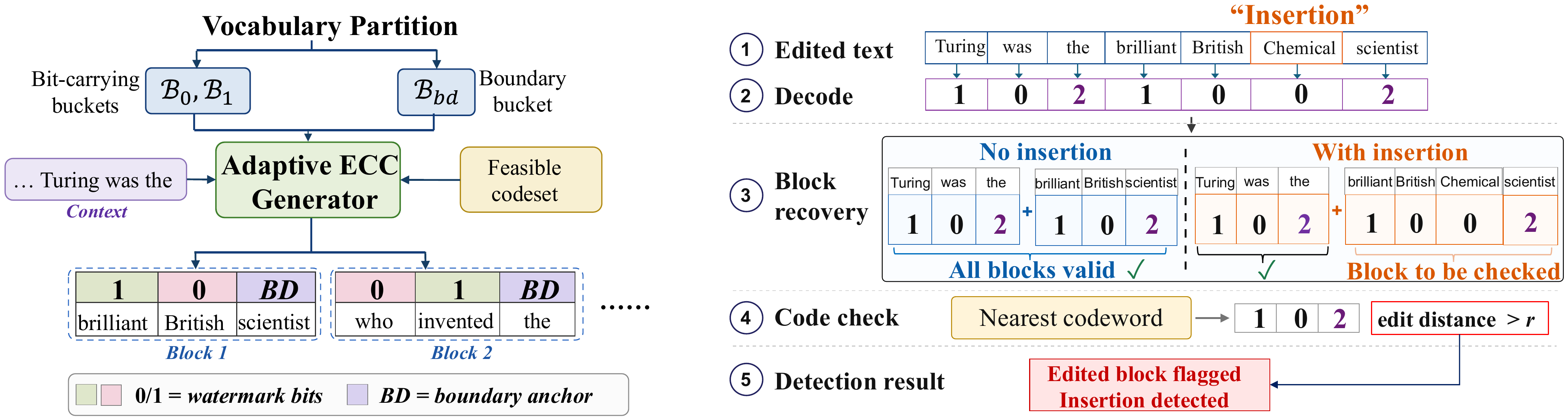}
    \caption{
    {\it Left:} Overview of the \algname{} watermark embedding procedure. The tokenizer vocabulary is partitioned into two buckets $\mathcal{B}_0,\mathcal{B}_1$ and a disjoint boundary-anchor bucket $\mathcal{B}_{\mathrm{bd}}$.
    The feasible codeword set $\mathcal{C}=\mathcal{V}_a(n)\cap\mathcal{H}$ combines VT and Hamming constraints.
    During \algname{} generation, the adaptive generator tracks the current block prefix and restricts or biases the next-token distribution toward feasible bits, while boundary anchors close each structural block. {\it Right:} Overview of block-addressable integrity detection and candidate refinement using an insertion example. The edited text is first decoded into structural symbols, globally parsed into structural blocks, and then checked against the feasible codeword set $\mathcal{C}$. A block is flagged as suspicious when its minimum edit distance to the feasible set exceeds the tolerance threshold $r$.
    }
    \label{fig:watermark}
\end{figure*}

Given $y$ and the watermark key, the detector maps the observed tokens to the same structural alphabet and recovers a minimum-cost decomposition into candidate ECC blocks. A recovered block is flagged when its payload distance exceeds the tolerance $r$ or its boundary is inferred to have been deleted or substituted; each alarm is further refined to candidate payload positions, insertion gaps, or boundary sites. Figure~\ref{fig:watermark} summarizes this generation and detection pipeline.

\subsection{Error-Correcting Code Design}\label{sec:ecc}

\paragraph{VT Codes for Insertion/Deletion Detection.}

VT codes are classical single insertion--deletion correcting codes defined over binary sequences \citep{lcvenshtcin1966binary}.
For a fixed length $n$ and syndrome parameter $a \in \{0,1,\dots,n\}$, the VT code $\mathcal{V}_a(n) \subset \{0,1\}^n$ is defined as
\begin{equation}\label{eq:vt}
\mathcal{V}_a(n)
\triangleq
\left\{
x\in \{0,1\}^n 
:
\sum_{i=1}^n i x_i \equiv  a \hspace{-0.1in} \pmod{n+1}
\right\}.    
\end{equation}
A single insertion or deletion predictably changes the weighted-sum constraint, enabling syndrome-based detection and localization. We use this property to enforce block-level consistency under insertion/deletion edits.

\paragraph{Hamming Codes for Substitution Detection.}

Hamming codes use linear parity constraints to detect and correct substitution (bit-flip) errors \citep{hamming1950error}. Given a parity-check matrix $H$, the corresponding Hamming code of length $n$ is
\begin{equation}\label{eq:hamming}
\mathcal{H}
=
\left\{
x \in \{0,1\}^n
:
H x^\top = 0
\right\},
\end{equation}
where $H$ is a pre-specified parity-check matrix.
If a single bit is flipped, the resulting non-zero syndrome uniquely identifies the error location.
At the structural level, a bucket-changing substitution appears as a payload-bit flip without changing sequence length.
Hamming-type parity constraints therefore provide a complementary detection layer to VT codes.

\paragraph{Joint Codeword Design.} We combine the VT and Hamming constraints by constructing a feasible codeword space that simultaneously allows detection of insertion--deletion and substitution edits.
Let $\mathcal{V}_a(n)$ denote a VT code with syndrome $a$, and let $\mathcal{H}$ denote a Hamming code of the same length. Combining the constraints in Eqs.~\eqref{eq:vt} and \eqref{eq:hamming}, we define the feasible codeword set as 
$$
\mathcal{C}=\mathcal{V}_a(n)\cap\mathcal{H}.
$$

This intersection represents the set of binary sequences that are simultaneously constrained by the VT condition for insertion--deletion sensitivity and by the Hamming parity condition for substitution sensitivity.
The VT syndrome parameter $a$ is fixed in our experiments, and the Hamming constraint is applied using the standard parity-check matrix over the original coordinate order.

% \zw{Checked the implementation: there is no coordinate permutation $\Pi$ in the current code. The feasible set is constructed directly as $\mathcal{C}=\mathcal{V}_a(n)\cap\mathcal{H}$ using the standard Hamming parity-check matrix.}

\subsection{\algname{} Watermarking}\label{sec:ecc-watermark}

% \begin{figure}[t!]
%     \centering \includegraphics[width=0.8\linewidth]{Figures/watermark-ECC.pdf}
%     \caption{Overview of VT-based watermarking. Given a watermarking key, a random VT codeword ($\VT_0(3)$ used in this illustrative example) is sampled for each block and a synchronization string is generated. These signals are combined to form token-level tags, which bias the next-token distribution so that each token is sampled from the corresponding vocabulary subset.}
%     %\liyan{Preset this 8-bucket partition, but at each time $t$, we shuffle these 8 subsets based on the previous token. }
% \label{fig:watermark}
% \end{figure}

% \begin{figure}[t!]
%     \centering
%     \includegraphics[width=0.95\linewidth]{Figures/watermark_pipeline.png}
%     \caption{
%     Overview of the proposed ECC watermark embedding procedure.
%     The tokenizer vocabulary is partitioned into two payload buckets $\mathcal{B}_0,\mathcal{B}_1$ and a disjoint boundary-anchor bucket $\mathcal{B}_{\mathrm{bd}}$.
%     The feasible codeword set $\mathcal{C}=\mathcal{V}_a(n)\cap\mathcal{H}$ combines VT and Hamming constraints.
%     During generation, the adaptive ECC generator tracks the current block prefix and restricts or biases the next-token distribution toward feasible payload bits, while boundary anchors close each structural block. 
%     }
%     \label{fig:watermark}
% \end{figure}

The \algname{} watermark consists of two major components: a mixed-constraint
code design with two bit-carrying buckets and a small pool of boundary anchors.
At a high level, the language model generates ordinary tokens in its native
vocabulary space, while the watermarking layer tracks only a derived symbolic
sequence consisting of watermark bits and boundary markers.

\paragraph{Two Bit-Carrying Buckets.}
Recall that $\mathcal{V}$ denotes the tokenizer vocabulary. After excluding permanently
banned tokens and setting aside tokens reserved for boundaries, we partition the
remaining usable vocabulary $\mathcal{V}_{\mathrm{wm}}$ into two disjoint buckets
$\mathcal{B}_0$ and $\mathcal{B}_1$, which represent watermark bits $0$ and $1$, respectively.
We do not assign punctuation, numbers, or other special lexical categories to
separate buckets.
Instead, the generator can use such tokens naturally whenever they remain
available. This retains the usable lexical categories in the two payload buckets without further vocabulary fragmentation.

\paragraph{Boundary Anchors.}
To separate consecutive ECC blocks, we reserve a boundary-anchor pool $\mathcal{B}_{\mathrm{bd}}$, disjoint from the two bit-carrying buckets. These anchors mark block boundaries and are not interpreted as watermark bits. Our integrity-detection experiments use a compact pool of 150 model-specific anchors. Keeping this pool sparse reduces the probability that a non-key-aware insertion or substitution accidentally creates a boundary symbol, which could introduce a spurious split and disturb block alignment.

\paragraph{Watermarking Process.}
As illustrated in Figure~\ref{fig:watermark} (left), during the watermarking process, each closed watermark block consists of $n$ watermark-bit symbols followed by one boundary anchor.
During generation, each emitted token is mapped to either a watermark bit through $\mathcal{B}_0,\mathcal{B}_1$ or to a boundary anchor through $\mathcal{B}_{\mathrm{bd}}$.
For a partial block prefix, the generator restricts or biases the next-token distribution toward bit values that keep the prefix compatible with at least one codeword in $\mathcal{C}$.
Once $n$ watermark-bit symbols have been generated, the next structural symbol is forced or encouraged to be a boundary anchor, and this closes the current block.

\paragraph{Watermark Identifiability.} We also evaluate whether the complete text remains identifiable as watermarked. Because the observed structural sequence may not be aligned with the beginning of a block, we test every phase $s\in\{0,\ldots,n\}$, defined as the number of leading symbols skipped before partitioning the remaining sequence into consecutive blocks. For each phase, $D_s$ measures how well the resulting blocks match the watermark structure by combining payload distance to the nearest valid codeword, boundary-anchor mismatches, unmatched prefix and suffix symbols, and disagreement with the length-derived block estimate $\widehat J=\operatorname{round}(T'/(n+1))$. Appendix~\ref{app:global_score} gives the exact cost. With $D^*=\min_s D_s$, we define
\begin{equation}\label{eq:global-score}
S_{\mathrm{global}} \triangleq \frac{1}{1+D^*/\max(1,\widehat J)}.
\end{equation}
Here, higher $S_{\mathrm{global}}$ indicates stronger consistency with the keyed ECC watermark.

%I was wondering if there is a way to have a figure for the Watermark Identifiability?

\subsection{Pretrained-Model Realization and Distortion Reduction}\label{sec:distortion}

The ECC layer specifies valid structural sequences and how to decode them, but it does not prescribe a unique linguistic realization mechanism. To instantiate it on existing pretrained models without retraining, we map tokens to structural buckets and intervene in the next-token distribution. We implement three complementary mechanisms that support different operating points between structural adherence and linguistic flexibility.

\paragraph{Soft Watermarking via Logit Biasing.} Hard watermarking realizes each structural bit by sampling exclusively from its designated bucket. We also consider a soft variant that adds a logit bias $\delta$ toward tokens in the desired bucket while retaining probability mass for alternatives.
Formally, for a desired bucket $\mathcal{B}$, similar to \citet{aaronson2023watermarking}, we modify the next-token logits from the original $z_t$ to $\tilde z_t$ as
\begin{equation}
\label{eq:logit-bias}
\tilde{z}_t(v) = 
\begin{cases}
z_t(v) + \delta, & v \in \mathcal{B}, \\
z_t(v), & v \notin \mathcal{B}.
\end{cases}    
\end{equation}
This preserves probabilistic flexibility while statistically encouraging compliance with the structural bit.
The bias parameter $\delta$ governs the detectability–distortion trade-off; we use $\delta=20$ as an approximate-hard setting and $\delta\in\{2,5\}$ for soft watermarking.

\paragraph{Semantic-Balanced Bucket Splitting.}
Watermark bits are realized through the two payload buckets defined above.
Naive partitioning can cause certain buckets to lack \emph{semantically} appropriate tokens, leading to unnatural generations.
To mitigate this, we construct semantically balanced buckets.
Inspired by the semantic-balanced partitioning strategy of \citet{guo-etal-2024-context-aware}, we group semantically similar tokens and distribute each group across the structural buckets. For the compact-anchor protocol, we select 150 clean single-token words with light semantic content and bounded frequency in the long-form question answering (LFQA) corpus \citep{blagojevic2021lfqa}, then split the non-anchor tokens in each group approximately evenly between $\mathcal{B}_0$ and $\mathcal{B}_1$. As a generation-quality ablation, we allocate one eighth of the eligible vocabulary to $\mathcal{B}_{\mathrm{bd}}$, distribute boundary quotas in proportion to the semantic-group sizes, spread the selected anchors across within-group LFQA frequency ranks, and balance the remainder between the two payload buckets. This preserves semantic coverage at payload and boundary positions without changing the ECC block format or decoder.

\paragraph{Adaptive Codeword Completion.}
The joint feasible set $\mathcal{C}$ contains multiple codewords, so we adopt an adaptive
codeword selection mechanism during token generation. At each payload position, we
identify the codewords that match the bits generated so far and collect their
next bits. The generator then restricts or biases sampling toward the
corresponding payload buckets. When both bit values remain possible, the
model's token distribution can choose between them without any intervention; when only one remains,
the generation rule favors that value via Eq.~\eqref{eq:logit-bias}. Each sampled bit therefore narrows the
set of possible codeword completions.

With finite logit bias $\delta$, the model may still sample a bit that makes the current
prefix inconsistent with every codeword in $\mathcal{C}$. We then compare the
current prefix with the equal-length prefix of every codeword, retain those
with the fewest mismatched positions, and use their next bits to guide the
following step.

% Generation remains flexible as long as $\mathcal{C}(p)$ is nonempty.
% This realizes adaptive codeword completion: the model is only constrained to remain on some feasible codeword path, but the specific codeword is selected implicitly by the sequence of generated payload bits.

% \paragraph{Distortion–Flexibility Trade-off.}
Together, soft biasing preserves probabilistic support, semantic-balanced splitting preserves lexical diversity, and adaptive completion uses codebook redundancy to provide multiple valid token choices. The ablation study in Section~\ref{sec:numerical} quantifies the resulting improvement relative to non-adaptive constrained generation.

\subsection{Edit Detection}\label{sec:edit-detect}

The detection of post-generation edits is performed in two steps, as visualized in Figure~\ref{fig:watermark} (right). First, we decode the observed token sequence to recover the ECC blocks and their associated information bits. We then check whether the recovered bits are consistent with the code space $\mathcal C$.

\paragraph{Decoding.}

%\subsubsection{Boundary Recovery via Global Edit-Distance Decoding}

We first map each token to its bucket label, obtaining a structural sequence over $\{0,1,2\}$, where $0$ and $1$ denote watermark bits and $2$ denotes a boundary anchor. Under mixed post-generation edits, boundary anchors may be deleted or substituted to disrupt local alignment. Therefore, rather than relying on observed boundary symbols alone, we formulate boundary recovery as a global decoding problem that searches for the minimum-cost block decomposition of the entire structural sequence.

Let $y:=\{x'_k\}_{k=1}^{T'}$ denote the observed token sequence.
We consider all possible segmentations of $y$ into candidate blocks, together with assignments of codewords and boundary states.
Each block is associated with:
$(i)$ a bit-carrying segment, and
$(ii)$ a boundary state indicating whether the boundary is intact, deleted, or substituted. 
We define the total decoding cost as
\begin{equation}\label{eq:cost}
\mathcal{L}
\triangleq
\sum_{k}
\Big(
d_{\mathrm{edit}}(y_k, c_k)
+
c_{\mathrm{bd}}(k)
\Big),    
\end{equation}
where $y_k$ is the observed token subsequence assigned to block $k$, $c_k \in \mathcal{C}$ is a candidate codeword, $d_{\mathrm{edit}}$ is the edit distance between the observed subsequence and the codeword, and $c_{\mathrm{bd}}(k)$ is a boundary penalty defined as
\[
c_{\mathrm{bd}}(k)
=
\begin{cases}
0, & \text{boundary intact}, \\
1, & \text{boundary deleted or substituted}.
\end{cases}
\]
% \paragraph{Boundary States.}
% We explicitly model three boundary conditions:
% \begin{itemize}
%     \item \emph{Intact:} the boundary anchor $2$ is observed at the end of the block;
%     \item \emph{Deleted:} the boundary anchor is missing, causing adjacent blocks to merge;
%     \item \emph{Substituted:} the boundary anchor is replaced by a payload symbol ($0$ or $1$).
% \end{itemize}
% These cases are jointly considered during decoding, allowing the algorithm to explain boundary corruption as part of the global optimization.
Then we recover the blocks by {\it minimizing} the total decoding cost $\mathcal L$ in Eq.~\eqref{eq:cost} over all possible block segmentations, codeword assignments, and boundary states. We implement this optimization using dynamic programming, which efficiently enumerates candidate segmentations and returns the minimum-cost decomposition. 
This decoding procedure can thus be viewed as a form of minimum-cost sequence parsing under edit noise.
Boundary positions are treated as latent variables inferred through global consistency rather than explicitly detected.
For structural-sequence length $L$ and per-block search budget $e$, the memoized decoder has complexity $O(L e |\mathcal C| n(n+e))$. In our experiments, we set $n=7$, $e=3$, and $|\mathcal C|=10$ as fixed constants; decoding scales approximately linearly with sequence length; Appendix~\ref{app:decoder_complexity} reports practical runtime.

% Importantly, this procedure is not a greedy block-by-block decision.
% Instead, it performs sequence-level decoding, jointly determining block boundaries and payload assignments.
% This enables recovery from boundary deletions and substitutions that would otherwise cause cascading alignment errors under local heuristics.

\begin{remark} We note that block-boundary recovery does not require every anchor to remain intact. Theoretically, under a sparse independent edit model with per-token edit rate $\epsilon \ll 1$, the probability that two consecutive boundary anchors are both deleted is on the order of $\epsilon^2$, which is negligible for small $\epsilon$. Thus, an isolated missing boundary can typically be resolved using adjacent intact anchors and the global block-decomposition decoder in Eq.~\eqref{eq:cost}, limiting cascading misalignment.
\end{remark}

\begin{remark}[Implicit Block Count Determination]
Although the number of blocks is not explicitly specified, the decoding objective implicitly regularizes the segmentation.
Since each block is matched with a fixed-length codeword and boundary edits incur additional cost, segmentations with unrealistic block counts (either too many or too few) result in higher total edit distance.
As a result, the optimal solution naturally recovers a block count that approximately matches the ground truth. 
\end{remark}

\paragraph{Detection.} 

Given the decoded block decomposition, we analyze the parsed bit-carrying sequence block by block.
For each block $j=1,\ldots,J$, let $\hat c^{(j)}$ denote the bit sequence assigned to that block by the decoder. 
We compare each parsed block symbol $\hat c^{(j)}$ with the feasible codeword set $\mathcal C$ under edit distance.
For a threshold $r\geq 0$, block $j$ is structurally consistent when its boundary is intact and $\min_{c\in\mathcal C}d_{\mathrm{edit}}(\hat c^{(j)},c)\leq r$. Therefore, a larger payload distance or a deleted/substituted boundary may flag the block as suspicious.

% The threshold parameter $r$ is used only after decoding to decide whether a parsed block should be counted as suspicious.
Under hard watermarking and the approximate-hard setting, we use $r=0$ as an unedited block is expected to be consistent with $\mathcal C$, so any observed inconsistency indicates a possible edit. 
For finite soft-bias settings, model outputs may occasionally produce blocks that deviate from the VT--Hamming feasible set $\mathcal C$ even before any post-generation edit is applied.
Therefore, to reduce false alarms caused in soft-bias regimes, we use $r=2$ for $\delta=2$ and $r=1$ for $\delta=5$ in our experiments.

% \liyan{Zewei: since we have coverage results in Sec 4, it is necessary to talk about how we perform token-level localization if we did that?}

%\paragraph{Within-Block Candidate Refinement.}
In addition to block-level detection, the detector can return a {\it candidate set} of possible edit locations for each flagged block. We obtain this set from the union of minimum-edit-distance alignments between the parsed block and its nearest codewords in $\mathcal C$. The candidates may be payload positions, insertion gaps, or the boundary position. When several codewords or alignments are equally plausible, especially after multiple or length-changing edits, we keep the locations suggested by all of them. This refinement narrows inspection to positions flagged by the codeword alignments.

\section{Numerical Experiments}\label{sec:numerical}

% We first outline the experimental setup used to evaluate the proposed watermarking framework.
% Our evaluation focuses on two aspects: (i) robustness to mixed post-generation edits, and (ii) generation quality under structural constraints.

\paragraph{Model and Data.}
We evaluate Qwen3-8B, Mistral-7B-Instruct-v0.3, and OPT-125M using model-specific semantic vocabulary partitions. We generate watermarked answers for 256 independent English long-form question answering (LFQA) prompts \citep{blagojevic2021lfqa}. LFQA provides a natural setting for the attribution threat studied here: a watermarked answer to a concrete question can later be locally altered to spoof a factual claim, stance, certainty level, or source while retaining the surrounding generated content.
All main ECC experiments target 18 boundary-closed watermark blocks per answer.
The default integrity detection uses the compact 150-anchor partition. We evaluate the one-eighth boundary allocation separately as a generation-quality variant under the same prompts, block target, adaptive nearest-feasible generation, and sampling protocol.
Unless otherwise specified, experiments use CUDA inference with 4-bit quantization. For English LFQA generation, tokenizer special tokens, control-character tokens, and tokens containing non-ASCII characters are masked.

% \paragraph{Edit simulation.}
To evaluate edit detection performance, we apply synthetic post-generation edits to the generated text.
Specifically, we consider structural insertions, deletions, and payload-bit substitutions applied uniformly at random; the substitution operation flips the selected bucket bit.
We use edit rates $\{0.2,0.4,0.6,0.8\}$ to select affected blocks and sample between one and $k$ edit operations in each selected block, for $k\in\{1,2,3\}$.

\paragraph{Baselines.}
We include the KGW watermarking scheme~\citep{kirchenbauer2023watermark} as a standard token-level watermarking baseline. Since KGW is not designed for explicit edit localization, we adapt it to our block-wise edit-detection protocol by applying the same synthetic mixed-edit attacks and reporting block-level detection metrics. 
We further compare against two recent watermark-segment localization methods, Zhao-AOL~\citep{zhao2025efficiently} and WaterSeeker~\citep{pan2025waterseeker}. Since both methods operate on token-level watermark scores rather than defining standalone watermarking schemes, we instantiate them using the same KGW-style red-green watermark source. We generate KGW-watermarked continuations under the same prompts, length budget, and mixed-edit protocol, and partition them into fixed 8-token blocks to match our ECC block length (with the boundary token). For each edited sequence, token-level KGW watermark-deficit scores are passed to Zhao-AOL or WaterSeeker, and the resulting token/span predictions are mapped to block-level edit predictions by overlap with edited block spans. We run all three KGW-based methods on each model with $\delta\in\{2,5,20\}$; Appendix~\ref{app:baseline_full} reports the complete comparison.

We also include two fixed-pattern combinatorial watermarking (CW) families evaluated in prior post-generation edit detection work~\citep{LLM-edit-detect2025}: the two-bucket alternating pattern (CW-AB) and four-bucket \texttt{ACADBCBD} pattern (CW-4). At each token, the original CW detector examines the $w$ overlapping windows that contain it ($w=2$ for CW-AB and $w=8$ for CW-4) and records the fraction that match the prescribed pattern. We flag the token only when none of these windows matches. Under the detector's strict threshold rule, this corresponds to $\tau_e=1/w$: $0.5$ for CW-AB and $0.125$ for CW-4. We use these thresholds at every watermark strength, and flag a block when any token in it is flagged. 

%Appendix~\ref{app:cw_calibration} reports the alternative clean-token Type-I-$0.1$ calibration and block-level ROC-AUC.

We also include the synchronization-assisted VT construction of \citet{deng2026detecting}, which is referred to as Sync-VT, as an ECC baseline for localized integrity checking. The construction is designed for insertion and deletion; Appendix~\ref{app:sync_ecc_baseline} separates its results by edit operation, including substitutions from the shared mixed-edit protocol.

\paragraph{Evaluation Metrics.}
We report block-level true positive rate (TPR), the fraction of edited blocks that are flagged, and false alarm rate (FAR), the fraction of unedited blocks that are flagged. Candidate coverage (Cov.) is the fraction of true edit events whose reference payload position, insertion gap, or boundary position appears in the candidate set returned for the corresponding block. Global verification reports ROC-AUC using the score in Eq.~\eqref{eq:global-score} against matched unwatermarked generations and human LFQA answers. Generation quality reports base-model PPL over saved continuations, either conditioned on the original prompt or scored without it.

% Perplexity is computed on the generated (clean) watermarked text using the same base language model without watermarking.
% Specifically, we compute corpus-level perplexity by aggregating token-level negative log-likelihood over all generated tokens:
% \[
% \mathrm{PPL} = \exp\left( \frac{1}{N} \sum_{t=1}^{N} -\log p(x_t \mid x_{<t}) \right),
% \]
% where the summation is taken over all tokens in all generated sequences.
% This corresponds to standard token-level averaged perplexity.

\paragraph{Cross-model Local Detection.}
Table~\ref{tab:three_model_local} shows that finite soft bias interacts with each model's next-token distribution: at $\delta\in\{2,5\}$, both clean code adherence and the TPR--FAR operating point vary across models. Once the embedded structure is nearly complete at $\delta=20$, all three models reach TPR above $0.996$. Mistral and OPT keep FAR below $0.008$, while Qwen3 has a higher FAR of $0.075$ because its clean generations contain fewer exactly feasible blocks. Within each model, candidate coverage increases with watermark strength; across the three models, it ranges from \(0.510\)–\(0.613\) at \(\delta=2\) and from \(0.781\)–\(0.802\) at \(\delta=20\). At the approximate-hard setting, the same codebook and decoder achieve consistently high edit sensitivity across all three models. 

\begin{table}[ht!]
    \centering
    \caption{Three-model LFQA results, macro-averaged over four edit rates and three edit budgets. Clean is the mean number of initially feasible blocks among 18; Cov. is candidate coverage.}
    \label{tab:three_model_local}
    \small \setlength{\tabcolsep}{4pt}
    \begin{tabular}{llrrrr}
    \hline
        Model & $\delta$ & Clean & TPR $\uparrow$ & FAR $\downarrow$ & Cov. $\uparrow$ \\
    \hline
        \multirow{3}{*}{Qwen3-8B}
        & 2  & 1.78 & 0.2621 & 0.0839 & 0.5103 \\
        & 5  & 6.84 & 0.5217 & 0.2107 & 0.6106 \\
        & 20 & 16.71 & 0.9966 & 0.0754 & 0.7811 \\
    \hline
        \multirow{3}{*}{Mistral-7B}
        & 2  & 1.70 & 0.2630 & 0.0851 & 0.5108 \\
        & 5  & 5.34 & 0.5443 & 0.2119 & 0.5874 \\
        & 20 & 17.93 & 0.9978 & 0.0071 & 0.8009 \\
    \hline
        \multirow{3}{*}{OPT-125M}
        & 2  & 6.56 & 0.1843 & 0.0115 & 0.6134 \\
        & 5  & 15.26 & 0.3366 & 0.0162 & 0.7576 \\
        & 20 & 18.00 & 0.9980 & 0.0025 & 0.8023 \\
    \hline
    \end{tabular}
\end{table}

\paragraph{Block-level Detection Results.}

Table~\ref{tab:bias_sweep_baseline_compare} compares all methods under the same Qwen3/LFQA protocol. At $\delta=2$, \algname{} and WaterSeeker use relatively conservative operating points, reaching $0.2621/0.0839$ and $0.2898/0.1951$ TPR/FAR, respectively. Sync-VT, KGW, Zhao-AOL, and the two CW variants attain TPRs from $0.7173$ to $0.9626$, but with FARs from $0.6093$ to $1.0000$. At $\delta=5$, \algname{} increases to $0.5217/0.2107$, while WaterSeeker remains more conservative at $0.3132/0.1273$; the other baselines attain TPRs from $0.7560$ to $0.9619$ with FARs from $0.4963$ to $0.9964$. At $\delta=20$, \algname{} reaches $0.9966/0.0754$, compared with $0.7464/0.1029$ for Sync-VT, $0.7772/0.1003$ for KGW, $0.7848/0.1592$ for Zhao-AOL, $0.3438/0.0259$ for WaterSeeker, $0.3863/0.1093$ for CW-AB, and $0.8290/0.1378$ for CW-4. Across the three watermarking strengths, \algname{} retains measurable detection sensitivity under soft bias, with TPR increasing as the watermark strength rises. For Sync-VT, insertion/deletion TPR remains above $0.95$ at $\delta=20$, while substitution accounts for most of its recall gap (Appendix~\ref{app:sync_ecc_baseline}).

\begin{table}[ht!]
    \centering
    \caption{Qwen3/LFQA block detection across watermark strengths, macro-averaged over four edit rates and $k\in\{1,2,3\}$.} \label{tab:bias_sweep_baseline_compare}
    \small   
    \vspace{-0.1in}\renewcommand{\arraystretch}{1.03}
    \begin{tabular*}{\textwidth}{@{\extracolsep{\fill}}lrrrrrr@{}}
    \hline
        & \multicolumn{2}{c}{$\delta=2$} & \multicolumn{2}{c}{$\delta=5$} & \multicolumn{2}{c}{$\delta=20$} \\
        \cline{2-7}
        Method & TPR $\uparrow$ & FAR $\downarrow$ & TPR $\uparrow$ & FAR $\downarrow$ & TPR $\uparrow$ & FAR $\downarrow$ \\
    \hline
        \textbf{\algname{}} & \text{0.2621} & \text{0.0839} & \text{0.5217} & \text{0.2107} & \text{0.9966} & \text{0.0754} \\
        Sync-VT & 0.8880 & 0.8880 & 0.8828 & 0.8793 & 0.7464 & 0.1029 \\
        KGW & 0.9558 & 0.8909 & 0.8774 & 0.5672 & 0.7772 & 0.1003 \\
        Zhao-AOL & 0.7173 & 0.6093 & 0.7560 & 0.4963 & 0.7848 & 0.1592 \\
        WaterSeeker & 0.2898 & 0.1951 & 0.3132 & 0.1273 & 0.3438 & 0.0259 \\
        CW-AB & 0.8025 & 0.8051 & 0.7703 & 0.7754 & 0.3863 & 0.1093 \\
        CW-4 & 0.9626 & 1.0000 & 0.9619 & 0.9964 & 0.8290 & 0.1378 \\
    \hline
    \end{tabular*}
\end{table}

\paragraph{LLM-guided Sparse Edits.}

Beyond random synthetic edits, we evaluate Qwen3-8B watermarked LFQA answers under question-aware LLM-guided editing. An instruction model proposes local edits under three benign motivations---grammar polishing, clarity improvement, and style softening---and three malicious motivations---claim distortion, stance shift, and source spoofing.
The instruction model outputs structured token-level edits, which are replayed on the saved generated token IDs to produce the edited sequence and exact block-level ground truth. We then map this sequence through the same vocabulary partition for detection.

Table~\ref{tab:llm_guided_edit_detection} reports the two bias settings separately. Each setting contains 72 edited examples, with 12 examples from each of the six motivations; the corresponding benign and malicious groups contain 36 examples each. At $\delta=5$, the detector achieves TPR $0.8261$, FAR $0.1814$, and candidate coverage $0.4169$. Increasing the bias to $\delta=20$ raises TPR to $0.9193$ and coverage to $0.5040$ while reducing FAR to $0.0921$. 
%Together with the perplexity results in Table~\ref{tab:boundary_eighth_quality}, these operating points show the trade-off between generation quality and local edit detection. 
Benign and malicious edits have similar detection sensitivity because the detector responds to structural change rather than intent.

\begin{table}[ht!]
    \centering
    \caption{
    Block detection and candidate coverage on the balanced Qwen3 LLM-guided edit sample. Edited blocks is the mean number per text.
    }    \label{tab:llm_guided_edit_detection}
    \small
    \setlength{\tabcolsep}{4pt}
    \begin{tabular}{cc|c|cccc}
    \hline
        $\delta$ & Intent & Texts & Edited blocks & TPR $\uparrow$ & FAR $\downarrow$ & Cov. $\uparrow$ \\
    \hline
        \multirow{3}{*}{5} & Overall & 72 & 5.75 & 0.8261 & 0.1814 & 0.4169 \\
        & Benign & 36 & 6.00 & 0.8380 & 0.2153 & 0.5088 \\
        & Malicious & 36 & 5.50 & 0.8131 & 0.1489 & 0.3422 \\
    \hline
        \multirow{3}{*}{20} & Overall & 72 & 5.33 & 0.9193 & 0.0921 & 0.5040 \\
        & Benign & 36 & 5.44 & 0.9337 & 0.0951 & 0.5520 \\
        & Malicious & 36 & 5.22 & 0.9043 & 0.0891 & 0.4608 \\
    \hline
    \end{tabular}
\end{table}

Table~\ref{tab:llm_guided_edit_detection} treats every block modified by the editor as edited, even when its final bucket sequence remains unchanged, as can occur when a token is replaced by another token in the same bucket. Appendix Table~\ref{tab:llm_edit_visibility} separately evaluates blocks whose bucket sequences do change. These structurally visible blocks account for $90.3\%$ of edited blocks at $\delta=5$ and $91.1\%$ at $\delta=20$. Among them, the detector flags $86.90\%$ at $\delta=5$ and all structurally visible blocks at $\delta=20$. The overall TPR in Table~\ref{tab:llm_guided_edit_detection} is lower because it also counts edited blocks whose bucket sequences remain unchanged.

\paragraph{Watermark Identifiability.}
Our primary evaluation concerns local edit detection, while the construction must also serve the conventional role of a watermark: distinguishing watermarked text from non-watermarked text. For each model and bias setting, we apply the global score in Eq.~\eqref{eq:global-score} to 256 watermarked answers and a combined negative set of 256 matched unwatermarked generations and 251 human LFQA answers. Across all three model families and watermark strengths, ROC-AUC is at least $0.9985$ (Table~\ref{tab:global_verification_auc}). The repeated ECC structure thus supports document-level watermark identification alongside block-level edit localization.

\begin{table}[ht!]
    \centering
    \caption{Global watermark-identifiability ROC-AUC under the compact-anchor protocol, using matched unwatermarked generations and human LFQA answers as negatives.}    \label{tab:global_verification_auc}
    \small
    \setlength{\tabcolsep}{7pt}
    \begin{tabular}{lrrr}
    \hline
        Model & $\delta=2$ & $\delta=5$ & $\delta=20$ \\
    \hline
        Qwen3-8B & 0.99847 & 0.99988 & 0.99998 \\
        Mistral-7B & 1.00000 & 0.99998 & 1.00000 \\
        OPT-125M & 0.99908 & 1.00000 & 0.99958 \\
    \hline
    \end{tabular}
\end{table}

\paragraph{Boundary Allocation and Perplexity.}
Each encoded block requires a boundary token, so the limited lexical choices in the compact 150-token boundary set may contribute substantially to generation cost. We test this effect with a quality-oriented variant that assigns one eighth of the eligible vocabulary to the boundary set and divides the remaining vocabulary evenly between the two payload buckets. Table~\ref{tab:boundary_eighth_quality} compares the two allocations under the same generation settings. The one-eighth allocation lowers conditional PPL at every model--bias setting, including reductions of $35.7$--$59.6\%$ at $\delta=20$.

\begin{table}[ht!]
    \centering
    \caption{Conditional perplexity (PPL) for compact-150 and boundary-$1/8$ anchor allocations (lower is better). ``Reduction'' reports the percentage decrease in PPL when replacing the default compact-150 allocation with boundary-$1/8$, providing an ablation study of how anchor allocation affects PPL.}    \label{tab:boundary_eighth_quality}
    \small   \setlength{\tabcolsep}{7pt}
    \begin{tabular}{lcrrr}
    \hline
        Model & $\delta$ & Compact-150 & Boundary-$1/8$ & Reduction \\
    \hline
        \multirow{3}{*}{Qwen3-8B} & 2  & 22.98 & 12.54 & 45.4\% \\
        & 5  & 32.27 & 15.43 & 52.2\% \\
        & 20 & 67.04 & 30.11 & 55.1\% \\
    \hline
        \multirow{3}{*}{Mistral-7B} & 2  & 11.20 & 8.27  & 26.1\% \\
        & 5  & 15.80 & 10.41 & 34.1\% \\
        & 20 & 44.68 & 18.04 & 59.6\% \\
    \hline
        \multirow{3}{*}{OPT-125M} & 2  & 18.89 & 14.71 & 22.1\% \\
        & 5  & 26.45 & 18.10 & 31.6\% \\
        & 20 & 30.20 & 19.43 & 35.7\% \\
    \hline
    \end{tabular}   
\end{table}

Matched unwatermarked conditional PPL is $1.96$, $2.07$, and $3.21$ for Qwen3, Mistral, and OPT, respectively, and all comparisons are within model. The ECC code and decoder are unchanged, so the reduction comes from additional lexical choices at boundary positions. A larger boundary set also increases the chance that an inserted or substituted token maps to a boundary, creating a spurious split or concealing a corrupted anchor. Boundary allocation therefore trades lower PPL against boundary distinctiveness and detection reliability.

\paragraph{Qualitative Generation Examples.}
Conditional PPL summarizes the distributional cost of watermark realization; the saved generations provide a direct view of the resulting text. Table~\ref{tab:qualitative_main} compares Qwen3-8B outputs for matched prompts at $\delta=2$ and $\delta=20$ under the compact-anchor integrity protocol. Across all three prompts, both operating points preserve the requested topic and central content. Appendix~\ref{app:source_generation_examples} provides five unmodified examples at each bias.

{\small    \renewcommand{\arraystretch}{1.03}
    \begin{longtable}{@{}p{0.09\textwidth}p{0.86\textwidth}@{}}
    \caption{Matched verbatim sentence prefixes from saved Qwen3-8B generations. Text is truncated only at sentence boundaries and is not rewritten.}\label{tab:qualitative_main} \\    
    \hline
    \multicolumn{2}{@{}p{0.95\textwidth}@{}}{\textbf{Prompt 1:} why do marathoners and triathletes tend to have small body frames?} \\*
    $\delta=2$ & Marathoners and triathletes frequently have smaller body frames due to the typical demands of their sports. \\*
    $\delta=20$ & Marathoners and triathlon athletes frequently have smaller body frames due to competitive modesty, which is an advantage since typical running efficiency reduces the amount waste carried across long distances. \\
    \hline
    \multicolumn{2}{@{}p{0.95\textwidth}@{}}{\textbf{Prompt 2:} what are magnet links? Context: I see them on lots of torrent sites.} \\*
    $\delta=2$ & Magnet links are a type of widely used hyperlink that allows users to locate nearly any file on the internet without requiring beforehand knowledge of where it is stored. \\*
    $\delta=20$ & Magnet links are URL-based addresses primarily used to locate and retrieve digital files across peer-to-peer network systems; they directly reference the content identified by hash values specifically generated for each item stored digitally along separate nodes. \\
    \hline
    \multicolumn{2}{@{}p{0.95\textwidth}@{}}{\textbf{Prompt 3:} why does the education system favours memory retention over imagination?} \\*
    $\delta=2$ & The education system often emphasizes memory retention primarily due to its focus on standardized testing alongside measurable outcomes. \\*
    $\delta=20$ & The education system tends to prioritize memory across various levels of learning due to measurable formal assessments and standardized testing, aiming mainly therefore at evaluating knowledge retention efficiently. \\
    \hline
    \end{longtable}
}

% \paragraph{Ablation: adaptive vs.\ non-adaptive generation.}
% We perform an ablation study to isolate the effect of adaptive codeword realization.
% In the adaptive variant, generation only enforces consistency with at least one feasible codeword continuation.
% In the non-adaptive variant, a feasible codeword is fixed in advance for each block.

% This comparison allows us to quantify the impact of adaptive generation on both robustness and generation quality.

% \section{Conclusion and Discussion}

% We proposed an ECC-based watermarking framework for detecting post-generation edits to watermarked LLM outputs. The method embeds local code constraints into the generated token sequence, and detects edits by decoding the observed structural sequence. Empirical results show that the proposed structure provides reliable block-level edit detection under mixed insertion, deletion, and substitution edits. These results suggest that error-correcting codes provide a useful mechanism for auditing local edits to watermarked LLM outputs.

\section{Conclusion}

We propose \algname{}, an ECC-based watermark that turns a document-level attribution signal into local evidence of post-generation edits. By combining VT and Hamming constraints with boundary anchors and global decoding, the watermark provides reliable block-level localization under sparse edits. Across Qwen3-8B, Mistral-7B-Instruct-v0.3, and OPT-125M, the approximate-hard setting achieves block-level TPR above $0.996$ and FAR at or below $0.0754$, while retaining strong global identifiability with ROC-AUC of at least $0.9985$. The decoder further narrows each block alarm to candidate token positions, insertion gaps, or boundary sites, giving reviewers a narrower region to inspect. Results on question-aware benign and spoofing edits show that this local evidence remains informative beyond randomly sampled token perturbations. We further explore two approaches to improving generation quality, each with a distinct trade-off. Soft watermarking reduces PPL with some loss in local detection reliability, whereas a larger boundary pool provides more lexical choices and reduces conditional PPL by $22.1$--$52.2\%$ under the soft settings, at the cost of a greater risk of boundary corruption. Together, these options provide flexible quality--reliability operating points for different application requirements.

% Our current study focuses on detecting sparse token-level insertions, deletions, and substitutions in watermarked LLM outputs. This setting captures an important class of local post-generation edits, but does not cover longer-span rewrites. Handling such cases may require more redundant code constructions, such as array VT codes, or layered watermark structures. Integrating multiple or layered watermarking schemes to jointly handle different types of edits is a promising direction for future work, with inherent trade-offs between robustness, redundancy, and text quality. In addition, our detector identifies structurally suspicious blocks but does not determine the semantic intent or severity of an edit, such as whether it is benign, malicious, or meaning-changing.

% Our empirical evaluation is limited in scope. The main experiments use one open-source autoregressive model and Wikipedia-style prompts, and most edits are synthetically simulated mixed-type token edits. We include a small-scale LLM-guided editing study to test more semantically motivated edits, but larger-scale evaluations across additional open-source models, domains, and naturally occurring or human-written post-generation edits are left as future work. We expect the approach to be broadly applicable to autoregressive models because the mechanism operates through token-level vocabulary partitioning and decoding, but validating this generality remains future work.

\bibliography{myreferences}
\bibliographystyle{plainnat}

\appendix

\section{Additional Numerical Results and Details}

\subsection{Global Watermark Score}
\label{app:global_score}

For phase $s$, let $J_s=\lfloor(T'-s)/(n+1)\rfloor$ be the number of complete blocks after skipping the first $s$ structural symbols. Let $p_j^{(s)}$ contain the first $n$ symbols of block $j$, and let $b_j^{(s)}$ be its final, expected boundary symbol. We define the following heuristic phase cost:
\[
\begin{split}
D_s ={}& \sum_{j=1}^{J_s}
\left[
\min_{c\in\mathcal C} d_H\!\left(p_j^{(s)},c\right)
+ \mathbf 1\!\left\{b_j^{(s)}\neq 2\right\}
\right] +s+\left[T'-s-J_s(n+1)\right]
+\left|J_s-\widehat J\right|.
\end{split}
\]
The first term measures payload disagreement with the nearest feasible codeword and boundary mismatch within each complete block. The next two terms count unmatched symbols before the first complete block and after the last one. The final term penalizes disagreement between the recovered block count $J_s$ and the length-based estimate $\widehat J$. The verifier uses the minimum cost across phases in Eq.~\eqref{eq:global-score}.

\subsection{Decoder Complexity and Practical Runtime}
\label{app:decoder_complexity}

The decoder uses a memoized dynamic program over positions in the observed structural sequence. At each position, it considers $2e+1$ candidate payload lengths from $n-e$ to $n+e$ and compares each candidate segment with every codeword in $\mathcal C$. A Levenshtein comparison costs $O(n(n+e))$, giving total complexity $O\!\left(L e |\mathcal C| n(n+e)\right)$,
where $L$ is the structural-sequence length. With $n=7$, $e=3$, and $|\mathcal C|=10$ fixed, the practical scaling is approximately linear in $L$. On an Intel i7-13650HX, decoding an 18-block sequence requires $31.6$ ms for clean text and $61.0$ ms under the strongest evaluated attack, including candidate-location extraction.

\subsection{Additional Results}

\subsubsection{Synchronization-Assisted VT Baseline}
\label{app:sync_ecc_baseline}

\begin{table}[H]
    \centering
    \caption{Sync-VT on Qwen3/LFQA, macro-averaged over four edit rates and $k\in\{1,2,3\}$. $^\dagger$The original construction addresses insertion and deletion; substitution results are reported for completeness under the shared mixed-edit protocol.}
    \label{tab:sync_ecc_attack_breakdown}
    \small
    \setlength{\tabcolsep}{5pt}
    \begin{tabular}{llrrr}
    \hline
        $\delta$ & Attack & TPR $\uparrow$ & FAR $\downarrow$ & Cov. $\uparrow$ \\
    \hline
        \multirow{3}{*}{2}
        & Insertion & 0.8890 & 0.8881 & 0.2994 \\
        & Deletion & 0.8878 & 0.8892 & 0.4149 \\
        & Substitution$^\dagger$ & 0.8872 & 0.8868 & 0.0000 \\
    \hline
        \multirow{3}{*}{5}
        & Insertion & 0.8828 & 0.8776 & 0.2983 \\
        & Deletion & 0.8885 & 0.8790 & 0.4176 \\
        & Substitution$^\dagger$ & 0.8771 & 0.8814 & 0.0000 \\
    \hline
        \multirow{3}{*}{20}
        & Insertion & 0.9581 & 0.0809 & 0.6463 \\
        & Deletion & 0.9515 & 0.0975 & 0.6761 \\
        & Substitution$^\dagger$ & 0.3297 & 0.1305 & 0.0000 \\
    \hline 
    \end{tabular}
\end{table}

\subsubsection{Full KGW-Based Baseline Comparison}
\label{app:baseline_full}

\begin{table}[H]
\caption{Full LFQA sweep for \algname{} and KGW-based baselines. Each entry is macro-averaged over four edit rates and $k\in\{1,2,3\}$.}
    \centering
    \small
    \setlength{\tabcolsep}{2.4pt}
    \begin{tabular}{ll|cc|cc|cc}
    \hline
        \multirow{2}{*}{Model} & \multirow{2}{*}{Method}
        & \multicolumn{2}{c|}{$\delta=2$}
        & \multicolumn{2}{c|}{$\delta=5$}
        & \multicolumn{2}{c}{$\delta=20$} \\
        & & TPR & FAR & TPR & FAR & TPR & FAR \\
    \hline
        \multirow{4}{*}{Qwen3-8B}
        & \textbf{\algname{}} & 0.2621 & 0.0839 & 0.5217 & 0.2107 & 0.9966 & 0.0754 \\
        & KGW & 0.9558 & 0.8909 & 0.8774 & 0.5672 & 0.7772 & 0.1003 \\
        & Zhao-AOL & 0.7173 & 0.6093 & 0.7560 & 0.4963 & 0.7848 & 0.1592 \\
        & WaterSeeker & 0.2898 & 0.1951 & 0.3132 & 0.1273 & 0.3438 & 0.0259 \\
    \hline
        \multirow{4}{*}{Mistral-7B}
        & \textbf{\algname{}} & 0.2630 & 0.0851 & 0.5443 & 0.2119 & 0.9978 & 0.0071 \\
        & KGW & 0.9642 & 0.9254 & 0.8913 & 0.6416 & 0.7833 & 0.1022 \\
        & Zhao-AOL & 0.7172 & 0.6248 & 0.7645 & 0.5382 & 0.7863 & 0.1615 \\
        & WaterSeeker & 0.2929 & 0.2149 & 0.3081 & 0.1360 & 0.3484 & 0.0249 \\
    \hline
        \multirow{4}{*}{OPT-125M}
        & \textbf{\algname{}} & 0.1843 & 0.0115 & 0.3366 & 0.0162 & 0.9980 & 0.0025 \\
        & KGW & 0.8886 & 0.6144 & 0.7939 & 0.1868 & 0.7780 & 0.1054 \\
        & Zhao-AOL & 0.7942 & 0.5410 & 0.7904 & 0.2312 & 0.7850 & 0.1596 \\
        & WaterSeeker & 0.3476 & 0.1150 & 0.3488 & 0.0358 & 0.3507 & 0.0237 \\
    \hline
    \end{tabular}
\end{table}

\subsubsection{Single-Edit Breakdown}

Table~\ref{tab:single_edit_block_adaptive} reports Qwen3 block-level TPR and FAR for $k=1$, with one uniformly sampled insertion, deletion, or substitution per affected block. \algname{} maintains near-perfect TPR across edit rates, whereas KGW and AOL have lower recall with higher FAR and WaterSeeker is conservative but low-recall.

\begin{table}[H]
    \centering
    \caption{
    Block-level edit detection under the single-edit setting ($k=1$), where each affected block receives one mixed edit sampled from substitution, insertion, and deletion.
    AOL denotes KGW+AOL, and WS denotes KGW+WaterSeeker.
    }    \label{tab:single_edit_block_adaptive}
    \vspace{-0.05in}
    \small
    \setlength{\tabcolsep}{6pt}
    \renewcommand{\arraystretch}{1}
    \begin{tabular}{c|l|cc}
    \hline
        Edit Rate & Method & TPR $\uparrow$ & FAR $\downarrow$ \\
    \hline
        \multirow{4}{*}{0.2}
        & \algname{} & {0.9989} & 0.0730 \\
        & KGW & 0.6922 & 0.0350 \\
        & Zhao-AOL & 0.7983 & 0.0776 \\
        & WaterSeeker  & 0.4568 & 0.0232 \\
    \hline
        \multirow{4}{*}{0.4}
        & \algname{} & {0.9973} & 0.0747 \\
        & KGW & 0.7052 & 0.0571 \\
        & Zhao-AOL & 0.7737 & 0.1132 \\
        & WaterSeeker  & 0.3365 & 0.0284 \\
    \hline
        \multirow{4}{*}{0.6}
        & \algname{} & {0.9968} & 0.0689 \\
        & KGW & 0.7246 & 0.0854 \\
        & Zhao-AOL & 0.7702 & 0.1522 \\
        & WaterSeeker  & 0.2914 & 0.0279 \\
    \hline
        \multirow{4}{*}{0.8}
        & \algname{} & {0.9973} & 0.0715 \\
        & KGW & 0.7244 & 0.1383 \\
        & Zhao-AOL & 0.7433 & 0.1933 \\
        & WaterSeeker  & 0.2572 & 0.0400 \\
    \hline
    \end{tabular}
\end{table}

\subsection{Source Text Generation for Question-Aware Editing}
\label{app:source_generation_examples}

We construct source continuations from 256 independent English LFQA prompts. Each prompt is a long-form question, and the Qwen3-8B continuation is a watermarked answer that can later be locally edited. This setup is well matched to malicious-edit evaluation: an editor can spoof how the answer addresses the question by changing certainty, attribution, numerical claims, or definitional boundaries. We use chat-template prompting with thinking disabled and adaptive \algname{} generation.

The following examples use the same five LFQA prompts at each $\delta\in\{2,5,20\}$ so that the effect of watermark strength can be inspected directly. Every response is generated text from the saved Qwen3-8B artifacts. We normalize whitespace for typesetting and truncate only at a sentence boundary; no displayed words are rewritten or selectively removed.

\input{qwen3_lfqa_examples.tex}

\subsection{LLM-Guided Editing and Downstream Risk Assessment}
\label{app:llm_guided_editing}

This appendix describes the LLM-guided editing experiment in more detail.
% \paragraph{Sparse LLM-Guided Editing Protocol.}
% We construct a controlled sparse editing setting using an instruction-tuned LLM as an edit-policy generator.
We start from the Qwen3-8B LFQA generations at $\delta=5$ and $\delta=20$.
For each selected continuation, the editor model is given the source prompt, the generated suffix, and an indexed edit-unit table.
Each edit unit contains an approximate token surface, its associated ECC block ID, bit index, and bucket ID.
The editor is also given one of six editing motivations.
We group 
\texttt{grammar\_polish}, \texttt{clarity\_improvement}, and \texttt{style\_softening}
as benign intents, and 
\texttt{claim\_distortion}, \texttt{stance\_shift}, and \texttt{source\_spoofing}
as malicious or adversarial intents.

The editor does not directly rewrite the full text. Instead, it outputs a JSON object containing local edit instructions, 
%\texttt{substitute}, \texttt{delete}, and \texttt{insert}. Each instruction 
which specifies a text-unit index or insertion gap, optional original text, new content when applicable, and a short reason.
The program validates the JSON, tokenizes the proposed new content, and applies the instructions to the saved generated token IDs. It then maps the complete edited token sequence through the same vocabulary partition and runs the detector on the resulting structural sequence. Editor instructions and ground-truth locations are used only for evaluation and are not supplied to the detector. This design gives exact block-level ground truth while still allowing the edit choices to be guided by semantic motivations.
The editor is asked to produce between six and ten local instructions, but chooses their operations and target locations. No detector output or evaluation metric is supplied to the editor.

\begin{table}[H]
    \centering
    \caption{Structural visibility in the balanced LLM-guided sample. A block is visible when its final bucket sequence differs from the original. Visible TPR is the fraction of these visible edited blocks that are flagged.}
    \label{tab:llm_edit_visibility}
    \small
    \setlength{\tabcolsep}{7pt}
    \begin{tabular}{c|rrr|r}
    \hline
        $\delta$ & Edited blocks & Visible blocks & Visibility & Visible TPR \\
    \hline
        5  & 414 & 374 & 0.9034 & 0.8690 \\
    \hline
        20 & 384 & 350 & 0.9115 & 1.0000 \\
    \hline
    \end{tabular}
\end{table}

Table~\ref{tab:llm_editor_by_motivation_app} reports the balanced sample by motivation and bias. Each stratum contains 216 evaluated blocks; the edited and clean counts provide the denominators for TPR and FAR, respectively. Candidate coverage is consistently lower than block TPR, reflecting the greater difficulty of within-block refinement.

\begin{table}[ht!]
    \centering
    \caption{
    Qwen3 LLM-guided edit results by motivation. E/C gives the numbers of edited and clean blocks. The first three motivations are benign and the last three are malicious.
    }
    \label{tab:llm_editor_by_motivation_app}
    \small
    \setlength{\tabcolsep}{4pt}
    \renewcommand{\arraystretch}{1.08}
    \begin{tabular}{cc|c|ccc}
    \hline
        $\delta$ & Motivation & Blocks (E/C) & TPR $\uparrow$ & FAR $\downarrow$ & Cov. $\uparrow$ \\
    \hline
        \multirow{6}{*}{5}
        & grammar\_polish & 72/144 & 0.8056 & 0.2292 & 0.5931 \\
        & clarity\_improvement & 66/150 & 0.8788 & 0.2000 & 0.4522 \\
        & style\_softening & 78/138 & 0.8333 & 0.2174 & 0.4870 \\
        & claim\_distortion & 65/151 & 0.8154 & 0.1192 & 0.3758 \\
        & stance\_shift & 66/150 & 0.7879 & 0.1533 & 0.3204 \\
        & source\_spoofing & 67/149 & 0.8358 & 0.1745 & 0.3368 \\
    \hline
        \multirow{6}{*}{20}
        & grammar\_polish & 75/141 & 0.8933 & 0.1348 & 0.5212 \\
        & clarity\_improvement & 60/156 & 0.9667 & 0.0897 & 0.5617 \\
        & style\_softening & 61/155 & 0.9508 & 0.0645 & 0.5764 \\
        & claim\_distortion & 55/161 & 0.9273 & 0.1118 & 0.4941 \\
        & stance\_shift & 65/151 & 0.8769 & 0.0530 & 0.3560 \\
        & source\_spoofing & 68/148 & 0.9118 & 0.1014 & 0.5494 \\
    \hline
    \end{tabular}
    
\end{table}

\end{document}

%% file: qwen3_lfqa_examples.tex
% Auto-generated from saved experiment artifacts; do not edit by hand.
% Whitespace is normalized and each response is truncated only at its first
% sentence boundary after 80 characters.
\subsubsection{$\delta=2$}

\noindent\fcolorbox{black!18}{black!2}{%
\parbox{\dimexpr\linewidth-2\fboxsep-2\fboxrule\relax}{%
\textbf{Example 1 (sample 10).}\\[-1pt]
\textbf{Prompt:} why do marathoners and triathletes tend to have small body frames?\\[2pt]
\textbf{Watermarked response:} Marathoners and triathletes frequently have smaller body frames due to the typical demands of their sports.%
}}
\par\vspace{6pt}

\noindent\fcolorbox{black!18}{black!2}{%
\parbox{\dimexpr\linewidth-2\fboxsep-2\fboxrule\relax}{%
\textbf{Example 2 (sample 188).}\\[-1pt]
\textbf{Prompt:} what are magnet links? Context: I see them on lots of torrent sites.\\[2pt]
\textbf{Watermarked response:} Magnet links are a type of widely used hyperlink that allows users to locate nearly any file on the internet without requiring beforehand knowledge of where it is stored.%
}}
\par\vspace{6pt}

\noindent\fcolorbox{black!18}{black!2}{%
\parbox{\dimexpr\linewidth-2\fboxsep-2\fboxrule\relax}{%
\textbf{Example 3 (sample 180).}\\[-1pt]
\textbf{Prompt:} why does the education system favours memory retention over imagination?\\[2pt]
\textbf{Watermarked response:} The education system often emphasizes memory retention primarily due to its focus on standardized testing alongside measurable outcomes.%
}}
\par\vspace{6pt}

\noindent\fcolorbox{black!18}{black!2}{%
\parbox{\dimexpr\linewidth-2\fboxsep-2\fboxrule\relax}{%
\textbf{Example 4 (sample 114).}\\[-1pt]
\textbf{Prompt:} how did the original mathematicians prove their formulas and theories, and to who?\\[2pt]
\textbf{Watermarked response:} Original mathematicians proved their formulas and formalized theories through rigorous logical reasoning, carefully structured arguments, and systematic exploration of previously established mathematical principles.%
}}
\par\vspace{6pt}

\noindent\fcolorbox{black!18}{black!2}{%
\parbox{\dimexpr\linewidth-2\fboxsep-2\fboxrule\relax}{%
\textbf{Example 5 (sample 190).}\\[-1pt]
\textbf{Prompt:} how come americans have large portion sizes and relatively cheap prices for their food?\\[2pt]
\textbf{Watermarked response:} The combination of large portion sizes and generally lower food prices in the United States largely stems from economic factors, industrial practices across the food industry, and consumer behavior alongside government policies.%
}}
\par\vspace{6pt}

\subsubsection{$\delta=5$}

\noindent\fcolorbox{black!18}{black!2}{%
\parbox{\dimexpr\linewidth-2\fboxsep-2\fboxrule\relax}{%
\textbf{Example 1 (sample 10).}\\[-1pt]
\textbf{Prompt:} why do marathoners and triathletes tend to have small body frames?\\[2pt]
\textbf{Watermarked response:} Marathoners and triathletes frequently have smaller body frames due to the typical demands of their sports.%
}}
\par\vspace{6pt}

\noindent\fcolorbox{black!18}{black!2}{%
\parbox{\dimexpr\linewidth-2\fboxsep-2\fboxrule\relax}{%
\textbf{Example 2 (sample 188).}\\[-1pt]
\textbf{Prompt:} what are magnet links? Context: I see them on lots of torrent sites.\\[2pt]
\textbf{Watermarked response:} Magnet links are a type of widely used hyperlink that allows users to locate nearly any file on the internet without requiring beforehand knowledge of where it is stored.%
}}
\par\vspace{6pt}

\noindent\fcolorbox{black!18}{black!2}{%
\parbox{\dimexpr\linewidth-2\fboxsep-2\fboxrule\relax}{%
\textbf{Example 3 (sample 180).}\\[-1pt]
\textbf{Prompt:} why does the education system favours memory retention over imagination?\\[2pt]
\textbf{Watermarked response:} The education system often emphasizes memory retention primarily due to its focus on standardized assessments alongside curriculum requirements.%
}}
\par\vspace{6pt}

\noindent\fcolorbox{black!18}{black!2}{%
\parbox{\dimexpr\linewidth-2\fboxsep-2\fboxrule\relax}{%
\textbf{Example 4 (sample 114).}\\[-1pt]
\textbf{Prompt:} how did the original mathematicians prove their formulas and theories, and to who?\\[2pt]
\textbf{Watermarked response:} Original mathematicians, such as Eu largely through logical deduction and geometric reasoning.early Greek mathematician Euclid, developed formal proofs by starting with axioms and gradually deriving theorems through deductive formal logic.%
}}
\par\vspace{6pt}

\noindent\fcolorbox{black!18}{black!2}{%
\parbox{\dimexpr\linewidth-2\fboxsep-2\fboxrule\relax}{%
\textbf{Example 5 (sample 190).}\\[-1pt]
\textbf{Prompt:} how come americans have large portion sizes and relatively cheap prices for their food?\\[2pt]
\textbf{Watermarked response:} The combination of large portion sizes and generally lower food prices in the United States largely stems from economic factors, industrial agriculture primarily focused on efficiency rather than quality or modest portions.%
}}
\par\vspace{6pt}

\subsubsection{$\delta=20$}

\noindent\fcolorbox{black!18}{black!2}{%
\parbox{\dimexpr\linewidth-2\fboxsep-2\fboxrule\relax}{%
\textbf{Example 1 (sample 10).}\\[-1pt]
\textbf{Prompt:} why do marathoners and triathletes tend to have small body frames?\\[2pt]
\textbf{Watermarked response:} Marathoners and triathlon athletes frequently have smaller body frames due to competitive modesty, which is an advantage since typical running efficiency reduces the amount waste carried across long distances.%
}}
\par\vspace{6pt}

\noindent\fcolorbox{black!18}{black!2}{%
\parbox{\dimexpr\linewidth-2\fboxsep-2\fboxrule\relax}{%
\textbf{Example 2 (sample 188).}\\[-1pt]
\textbf{Prompt:} what are magnet links? Context: I see them on lots of torrent sites.\\[2pt]
\textbf{Watermarked response:} Magnet links are URL-based addresses primarily used to locate and retrieve digital files across peer-to-peer network systems; they directly reference the content identified by hash values specifically generated for each item stored digitally along separate nodes.%
}}
\par\vspace{6pt}

\noindent\fcolorbox{black!18}{black!2}{%
\parbox{\dimexpr\linewidth-2\fboxsep-2\fboxrule\relax}{%
\textbf{Example 3 (sample 180).}\\[-1pt]
\textbf{Prompt:} why does the education system favours memory retention over imagination?\\[2pt]
\textbf{Watermarked response:} The education system tends to prioritize memory across various levels of learning due to measurable formal assessments and standardized testing, aiming mainly therefore at evaluating knowledge retention efficiently.%
}}
\par\vspace{6pt}

\noindent\fcolorbox{black!18}{black!2}{%
\parbox{\dimexpr\linewidth-2\fboxsep-2\fboxrule\relax}{%
\textbf{Example 4 (sample 114).}\\[-1pt]
\textbf{Prompt:} how did the original mathematicians prove their formulas and theories, and to who?\\[2pt]
\textbf{Watermarked response:} Original mathematicians often used logical deduction alongside empirical observation to validate formulas and theories across various mathematical domains throughout history.%
}}
\par\vspace{6pt}

\noindent\fcolorbox{black!18}{black!2}{%
\parbox{\dimexpr\linewidth-2\fboxsep-2\fboxrule\relax}{%
\textbf{Example 5 (sample 190).}\\[-1pt]
\textbf{Prompt:} how come americans have large portion sizes and relatively cheap prices for their food?\\[2pt]
\textbf{Watermarked response:} The combination of large portion sizes and generally lower food prices in the United States largely stems from economic incentives aimed at increasingoverall consumption.%
}}
\par\vspace{6pt}